# Stability of numerous novel potassium chlorides at high pressure

Weiwei Zhang[1, 2, *], Artem R. Oganov[2-4, *]

**K-Cl is a simple system displaying all four main types of bonding, as it contains (i) metallic potassium, (ii) elemental chlorine made of covalently bonded $Cl_2$ molecules held together by van der Waals forces, and (iii) an archetypal ionic compound KCl. Well-known octet rule (and its special case, the charge balance rule, assigning classical charges of "+1" to K and "-1" to Cl) predicts that no compounds other than KCl are possible. However, our quantum-mechanical variable-composition evolutionary simulations predict new stable compounds. The phase diagram of the K-Cl system turned out to be extremely complicated, featuring new thermodynamically compounds $K_3Cl$, $K_2Cl$, $K_3Cl_2$, $K_4Cl_3$, $K_5Cl_4$, $K_3Cl_5$, $KCl_3$, $KCl_7$. Of particular interest are 2D-metallic homologs $K_{n+1}Cl_n$, the presence of positively charged Cl atoms in $KCl_7$, and the predicted stability of $KCl_3$ already at the atmospheric pressure.**

Recent *ab initio* calculations predicted the formation of unexpected novel high-pressure compounds in several simple systems, such as Li-H[1-2,] Na-H[3], K-H[4], Mg-O[5], and Na-Cl[6]. Two of these systems were subsequently explored experimentally: while so far the predictions have not been verified for Li-H[1], for Na-Cl the predicted compounds $NaCl_3$ and $Na_3Cl$ have been confirmed experimentally[6], revealing one of the most dramatic changes of chemistry under pressure. Here we study a closely related system, K-Cl, and find even richer chemistry and new phenomena.

The only known potassium chloride, KCl, has been extensively studied under pressure, both experimentally[7-9] and using *ab initio* simulations[10-12]. Two crystal structures are known for KCl: the rocksalt-type (B1) structure and cesium chloride-type (B2) structure, the latter becoming stable at ~2 GPa. The same transition happens in NaCl, but at a much higher pressure of 30 GPa [13-14]. This is consistent with the general tendency for phase transitions to occur at lower pressures for compounds of heavier elements – therefore, compounds similar to the exotic Na-Cl compounds[6], can be expected in the K-Cl system at lower pressures. This expectation is fulfilled only partly - the K-Cl system shows even more surprises than Na-Cl. Here we study the K-Cl system using variable-composition evolutionary structure prediction methodology USPEX[15-18], searching for the stable compounds and their corresponding crystal structures (see Methods). In each of these calculations, all possible chemical compositions were allowed with up to 16 atoms in the unit cell, and calculations were performed at pressures of 1 atm, 50 GPa, 100 GPa, 150 GPa, 200 GPa, 250 GPa and 300 GPa.

[1]Department of Applied Physics, China Agricultural University, Beijing, 100080, P.R.China. [2]Department of Geosciences, Center for Materials by Design, and Institute for Advanced Computational Science, State University of New York, Stony Brook, NY 11794-2100. [3]Moscow Institute of Physics and Technology, 9 Institutskiy Lane, Dolgoprudny city, Moscow Region 141700, Russia. [4]Northwestern Polytechnical University, Xi'an 710072, China
*These authors contributed equally to this work.

**Results and Discussion**

The pressure-composition phase diagram predicted in our calculations (Fig. 1, see also Supporting Online Materials) contains a surprisingly large number of new stable compounds are found. By thermodynamically stable we mean a compound which is more stable than any isochemical mixture of the elements or other compounds – this definition leads to the convex hull construction shown in Fig. 1b, c. The dynamical stability of the newly predicted phases was checked and confirmed by phonon calculations (see Supporting Online Material).

The phase diagram and crystal structures of the newly predicted compounds are remarkable. While we reproduce stability of KCl in the whole pressure range investigated here, many new stable compounds are predicted. The first surprise we find is $KCl_3$ is predicted to be stable already at 1 atm. Its low-pressure phase, stable from 0 to 34 GPa, has space group

*Pnma* and contains asymmetric trichloride $Cl_3^-$ molecular ions (Fig.2a). This compound has never been reported before, but there may be a reason: $Cl_3^-$ ions are unstable in the solution and high-temperature synthesis (e.g., from the melt) may require elevated pressures of a few GPa to suppress high entropy of $Cl_2$ fluid. This compound is a semiconductor with the DFT band gap of 2.44 eV at 20 GPa (more accurate GW calculations give 4.56 eV). This is the only non-metallic phase among all the newly predicted phases reported here. Bader analysis gives the charge configuration $K^{+0.8}[Cl^{-0.26}Cl^{-0.03}Cl^{-0.49}]^{-0.8}$, nearly the same as for *Pnma*-$NaCl_3$ [Ref. 6]. $[Cl_3]^-$ ion is an isoelectronic analogue of the well-known triiodide-ion $[I_3]^-$ (for example, compound $KI_3$ is well known), $Br_3^-$ and $ClICl^-$ ions, and can be also related to the known $[Li_3]^-$ [Ref. 19] and hypothetical $[H_3]^-$ [Ref. 2] ions. Considering the $Cl_3^-$ ion, we can see at least two possible interpretations. In a Zintl-like scheme, the middle Cl atom would have to be positively charged to form two covalent bonds. In the valence-shell electron pair repulsion model[20], the central Cl atom of the $Cl_3^-$ group has the $dsp^3$ hybridization, implying 5 electron pairs, a negative net charge, and violation of the octet rule. While the $I_3^-$ ion is traditionally described by the second model, for chlorine it is more problematic to populate vacant d-orbitals and the two schemes work simultaneously in resonance — explaining the nearly zero charge of this atom. At 34 GPa, this insulating state breaks down, and $KCl_3$ metallizes.

At 34 GPa, $KCl_3$ transforms into a $Cr_3Si$-type (also known as A15- type and related to the β-W structure type) metallic structure with space group $Pm3n$ — here, each K atom is surrounded by twelve Cl atoms that form an icosahedron (Fig. 2b). The same structure was found theoretically and experimentally for $NaCl_3$[6], but due to larger compressibility of K atoms, the difference in lattice parameters of $KCl_3$ and $NaCl_3$ (4.169 Å v 4.038 Å at 240 GPa) is smaller than what could be expected on the basis of the ionic radii (1.38 Å for $K^+$ and 1.02 Å for $Na^+$). Each unit cell of $KCl_3$-$Pm3m$ contains two K atoms and six Cl atoms. Substituting one K atom by a Cl atom, one obtains $KCl_7$ – a compound stable above 225 GPa, as found in our evolutionary searches. It has a cubic space group $Pm\bar{3}$ and is shown in Fig. 2c. Similar phases $NaCl_7$ and $NaCl_3$ have been predicted recently[6]. Very peculiarly, and consistently with $NaCl_7$[6], the substituting Cl atom, located in the center of $Cl_{12}$-icosahedron, has a slightly positive Bader charge of +0.06 (Table 1). The origin of this phenomenon is easy to trace to the Zintl model: to enable more than one Cl-Cl bond, the central Cl atom must be stripped of some of its valence electrons, thus attaining a positive charge. Note that $Cl^{+0.06}$ atoms are located in the centers of $Cl_{12}$ icosahedra formed by $Cl^{-0.09}$ atoms.

In the K-Cl system, in contrast with Na-Cl, there is yet another chlorine-rich phase, $K_3Cl_5$-$P\bar{4}m2$, which has a pseudocubic cell with 1 formula unit. The K atom in the center of the unit cell is surrounded by 4 K atoms and 10 Cl atoms, together forming a bicapped hexagonal antiprism (Fig. 2d). This structure is totally new; it is not stable in the Na-Cl system and has no known analogues in other systems. Fig. 3a shows its electronic structure. This is a poor metal with a deep pseudogap of width ~4.6 eV at 240 GPa, and the density of states at the Fermi level is less than 0.5 states/ (eV*cell). It is also quite clear that the largest maxima of the valence electron localization function (Fig. 3b) are around Cl atoms, and, in this and in other phases studied here, crystallographically inequivalent Cl atoms have very different ELF distributions − from spherical (around atoms with the most negative Bader charge, indicating a closed-shell configuration, but also around the positively charged Cl atom in KCl$_7$) to toroidal (around atoms with small negative charges).

We compare the total and atom-projected electronic densities of states of $P\bar{4}m2$-K$_3$Cl$_5$, $Pm3n$-KCl$_3$ and $Pm3$-KCl$_7$ in Fig. 3c, d, e. All these phases are poor metals with pronounced pseudogaps at the Fermi level, implying electronic stabilization. The main contribution at the Fermi level comes from chlorine atoms, and one can observe that different chlorine sites play very different roles – for example, in $P\bar{4}m2$-K$_3$Cl$_5$ only p-orbitals of Cl (4j) contribute at the Fermi level, and are thus responsible for its metallicity. Due to excess of chlorine atoms, which act as electron acceptors, $Pm3n$-KCl$_3$ has DOS similar to p-type semiconductors. The central, positively charged Cl atom, donating electrons to the system in $Pm3$-KCl$_7$, makes the DOS of this compound at the Fermi level much higher than that in $P\bar{4}m2$-K$_3$Cl$_5$ and $Pm3n$-KCl$_3$.

From Bader analysis (Table 1), we see that in K-rich (n<1) compounds, Cl has a nearly constant charge of ~-0.65, and K atoms adjust their charge to achieve electroneutrality. The opposite is found in Cl-rich KCl$_n$ compounds (n>1), where the Bader charge of K remains nearly constant, at about +0.6, while the charge of Cl is adapted to satisfy electroneutrality and becomes less negative with increasing Cl content (as discussed above, in KCl$_7$ one of chlorine atoms even has a positive charge). Somewhat higher Bader charges (about +0.8) were found[6] for Na atoms in $Pnma$-NaCl$_3$, $Pm3n$-NaCl$_3$ and $Pm3$-NaCl$_7$. This is consistent with our finding (Dong et al., in prep.) that under pressure K has higher electronegativity and lower reactivity than Na – opposite to ambient conditions, and due to the well-known s→d electronic transition in K atoms under pressure, making K a transition metal. Related to this is the observation that the depth of the convex hull (i.e. the enthalpy of formation of KCl or NaCl) in the K-Cl system (Fig. 1) changes from -2.9 eV/atom at 40 GPa to -1.5 eV/atom at 300 GPa, whereas for the Na-Cl system[6] it changes from -2.5 eV/atom at 40 GPa to -2.9 eV/atom at 300 GPa. Decreased reactivity of K and increasingly shallow K-Cl convex hull (Fig. 1) explain the variety of stable K-Cl compounds. While at normal conditions the electronegativity difference $\Delta\chi$(K-Cl) is 2.2, at 300 GPa it drops to mere 1.5 (Dong et al., in prep.) – similar to that in pairs Ti-O, Mo-O, Ti-N. Indeed, compounds with numerous stoichiometries, including "unusual" ones, are known in these systems – e.g., $Ti_3O$, $Ti_2O$, TiO, $Ti_2O_3$, $Ti_3O_5$, $Ti_4O_7$, $Ti_5O_9$, $TiO_2$, $Ti_2N$, TiN, $Mo_4O_{11}$, $Mo_5O_{14}$, $Mo_8O_{23}$, $Mo_9O_{26}$, $Mo_{14}O_{47}$, $MoO_2$, $MoO_3$, and many of these form homologous series with intermediate members featuring metal-metal bonds, just as we find in the K-Cl system.

In the studied pressure range, besides the known B1 and B2 phases, we find a new phase of KCl: $I4_1/amd$–KCl, stable above 201 GPa, shown in Fig.4a. This structure is a derivative of the fcc structure. Fig.4b shows $K_3Cl$, the other fcc-derived superstructure compound stable in the K-Cl system (above 149 GPa) – square planar layers with stoichiometry KCl alternate with similar layers of stoichiometry $K_2$ along the $c$-axis, leading to the total stoichiometry $K_3Cl$. These two compounds can be described as fcc-based homologs.

There is another interesting and surprisingly rich class of phases, $K_{n+1}Cl_n$ homologs ($n$=2, 3, 4 were found in our calculations, but we cannot exclude the possibility of even larger–$n$ homologs) based on the B2 structure and shown in Fig. 5. These have (2$n$+1) layers along the $c$-axis, with extra K-layer serving as an antiphase boundary between B2-structured domains. All these phases have the same space group $I4/mmm$, and similar distances between K and Cl. All of them are poor metals, due to the excess of electron-donating K atoms, analogous to n-type semiconductors (Fig. S2), and display a two-dimensional metallic character. It is rather surprising that phases with different *n* have rather different stability fields: e.g., $K_2Cl$ is stable at pressures above 56 GPa, whereas $K_5Cl_4$ is stable above 100 GPa. Interestingly, metallicity is observed only at the antiphase boundaries, whereas regions between them are insulating (Fig. 5e). These antiphase boundaries may be created as metastable growth defects also at lower pressures, with the promise of new electronic materials.

In summary, for a seemingly simple K-Cl system our calculations predict an extremely unusual behavior. Already at ambient pressure we predict stability of the new insulating compound $KCl_3$, which has not been observed before. As pressure increases, a surprisingly large number of thermodynamically stable phases become stable: (1) Cl-rich metallic phases ($KCl_7$, $K_3Cl_5$, and a metallic form of $KCl_3$) with high coordination numbers (12-14), (2) fcc-superstructures (insulating *I4$_1$/amd*-KCl and metallic $K_3Cl$), (3) layered B2-superstructures with compositions $K_{n+1}Cl_n$ (*n*=3,4,5) and two-dimensional electronic conductivity. What was considered as an ultimately simple chemical system, upon careful theoretical study turned out to be a very rich system with novel physics and chemistry. Revisiting other simple systems may result in the formulation of new chemical principles that could be used for the discovery of novel materials and phenomena.

**Methods**

Structure/composition predictions were done using the USPEX code[15-17] in the variable-composition mode[18]. The first generation of structures was produced randomly and the subsequent generations were obtained by applying heredity, transmutation, softmutation, and lattice mutation operations, with probabilities of 60%, 10%, 20% and 10%, respectively.

60% fittest non-identical structures of each generation were used to produce the next generation. 20% new random symmetric structures were also added in each generation. All structures were relaxed using density functional theory (DFT) calculations at the generalized gradient approximation level of theory, with the Perdew-Burke-Ernzerhof (PBE)[21] exchange-correlation functional, using the VASP code[22]. We used the all-electron projector augmented wave (PAW)[23] with K [$3s^23p^64s^1$], Cl [$2s^22p^4$] cores (core radii 2.20 a.u. and 1.50 a.u., respectively) and plane-wave basis sets with the 500 eV cutoff, and dense Monkhorst-Pack meshes with resolution better than $2\pi \times 0.05$Å$^{-1}$. We used the normalized enthalpy of formation as fitness and analyzed results of USPEX using the STM4 package[24]. Having identified the most stable compositions and structures, we relaxed them at pressures between 1 atm and 300 GPa using very accurate Brillouin zone sampling (Monkhorst-Pack meshes with resolution of better than $2\pi \times 0.03$Å$^{-1}$).

**Acknowledgements:**

We thank Xiao Dong for discussions and National Science Foundation (EAR-1114313, DMR-1231586) and DARPA (Grants No. W31P4Q1310005 and No. W31P4Q1210008) for financial support.

**Author contributions:**

A.R.O. designed the research. W.W.Z. and A.R.O. performed the calculations, interpreted data and wrote the paper.

**Additional information:**

Supplementary information is available.

**Competing financial interests**

The authors declare no competing financial interests.

**Table 1.** Structures of B1-KCl at 1 atm, B2-KCl and *Pnma*-KCl$_3$ at 20 GPa, A15-type (*Pm*3*n*) KCl$_3$, *Pm*3-KCl$_7$, $P\bar{4}m2$-K$_3$Cl$_5$, *I*4/*mmm*-K$_3$Cl and *I*4$_1$/*amd*-KCl at 240 GPa, and the corresponding atomic Bader charges (Q) and volumes (V).

**Figure 1. Stability of new potassium chlorides:** (a) Pressure-composition phase diagram of the K-Cl system. (b – c) Convex hull diagrams for the K-Cl system at selected pressures. Solid circles represent stable compounds; open circles - metastable ones.

**Figure 2. Crystal structures of** (a) *Pnma*-KCl$_3$ at 20 GPa (b) *Pm*3*n*-KCl$_3$ at 240 GPa (c) *Pm*3-KCl$_7$ at 240 GPa, (d) K$_3$Cl$_5$-$P\bar{4}m2$ at 240 GPa. Large gray spheres – K atoms, small green spheres – Cl atoms.

**Figure 3. Electronic structure:** (a) band structure and electronic density of states of $P\bar{4}m2$-K$_3$Cl$_5$ at 240 GPa, (b) electron localization function of $P\bar{4}m2$-K$_3$Cl$_5$ at 240 GPa with isosurface ELF=0.77. (c)total and atom-projected densities of states of $P\bar{4}m2$-K$_3$Cl$_5$, (d) total and atom-projected densities of states of *Pm*3*n*-KCl$_3$, (e) total and atom-projected densities of states of *Pm*3-KCl$_7$ at 240 GPa.

**Figure 4. Crystal structures of fcc-derived phases** (a) *I*4$_1$/*amd*-KCl (b) *I*4/*mmm*-K$_3$Cl. Large gray spheres – K atoms, small green spheres – Cl atoms.

**Figure 5. Crystal structures of K$_{n+1}$Cl$_n$ homologs** (a) *I*4/*mmm*-K$_2$Cl, (b) *I*4/*mmm*-K$_3$Cl$_2$, (c) *I*4/*mmm*-K$_4$Cl$_3$, (d) *I*4/*mmm*-K$_5$Cl$_4$ (e) Spatial distribution of electrons (shown by isosurfaces and density contours) at the Fermi level in *I*4/*mmm*-K$_5$Cl$_4$, showing clear 2D-metallic character. Large gray spheres – K atoms, small green spheres – Cl atoms.

**Table 1.** Structures of B1-KCl at 1 atm, B2-KCl and *Pnma*-KCl$_3$ at 20 GPa, A15-type (*Pm3n*) KCl$_3$, *Pm3*-KCl$_7$, $P\bar{4}m2$-K$_3$Cl$_5$, *I4/mmm*-K$_3$Cl and *I4$_1$/amd*-KCl at 240 GPa, and the corresponding atomic Bader charges (Q) and volumes (V).

|  | Lattice Parameters |  | X | y | Z | Q, \|e\| | V, Å$^3$ |
|---|---|---|---|---|---|---|---|
| B1-KCl | $a$ = 3.192Å | K(4b) | 0.500 | 0.500 | 0.500 | +0.843 | 22.875 |
|  |  | Cl(4a) | 0.000 | 0.000 | 0.000 | −0.843 | 42.165 |
| B2-KCl | $a$ = 3.350Å | K(1a) | 0.000 | 0.000 | 0.000 | +0.784 | 15.226 |
|  |  | Cl(1b) | 0.500 | 0.500 | 0.500 | −0.784 | 22.384 |
| *Pnma*-KCl$_3$ | $a$ = 8.149 Å<br>$b$ = 4.910 Å<br>$c$ = 7.276 Å | K(4c) | 0.166 | 0.250 | 0.471 | +0.785 | 14.70 |
|  |  | Cl(4c) | 0.922 | 0.250 | 0.763 | −0.263 | 19.35 |
|  |  | Cl(4c) | 0.618 | 0.250 | 0.543 | −0.033 | 18.33 |
|  |  | Cl(4c) | 0.845 | 0.250 | 0.341 | −0.489 | 20.39 |
| *Pm3n*-KCl$_3$ | $a$ = 4.169 Å | K(2a) | 0.000 | 0.000 | 0.000 | +0.519 | 7.91 |
|  |  | Cl(6c) | 0.000 | 0.500 | 0.250 | −0.173 | 9.43 |
| *Pm3*-KCl$_7$ | $a$ = 4.123 Å | K(1a) | 0.000 | 0.000 | 0.000 | +0.493 | 7.77 |
|  |  | Cl(1b) | 0.500 | 0.500 | 0.500 | +0.063 | 8.47 |
|  |  | Cl(6g) | 0.745 | 0.500 | 0.000 | −0.089 | 8.99 |
| $P\bar{4}m2$-K$_3$Cl$_5$ | $a$ = 4.136 Å<br>$c$ = 4.361 Å | K(1c) | 0.500 | 0.500 | 0.500 | +0.566 | 8.47 |
|  |  | K(2g) | 0.500 | 0.000 | 0.245 | +0.542 | 8.18 |
|  |  | Cl(1b) | 0.500 | 0.500 | 0.000 | −0.512 | 10.11 |
| | | Cl(4j) | 0.252 | 0.000 | 0.747 | -0.285 | 9.91 |
| $I4_1/amd$ -KCl | $a$ = 3.340Å $c$ = 6.873Å | K(4a) | 0.000 | 0.250 | 0.875 | +0.545 | 8.595 |
| | | Cl(4b) | 0.000 | 0.750 | 0.625 | -0.545 | 10.579 |
| $I4/mmm$-K$_3$Cl | $a$ = 3.365Å $c$ = 6.583Å | K(2b) | 0.500 | 0.500 | 0.000 | +0.288 | 8.818 |
| | | K(4d) | 0.500 | 0.000 | 0.250 | +0.249 | 8.863 |
| | | Cl(2a) | 0.000 | 0.000 | 0.000 | -0.786 | 10.726 |
| $I4/mmm$-K$_2$Cl | $a$= 2.749Å $c$ = 7.587Å | K(4e) | 0.000 | 0.000 | 0.330 | +0.344 | 9.19 |
| | | Cl(2a) | 0.000 | 0.000 | 0.000 | -0.688 | 11.30 |
| $I4/mmm$-K$_3$Cl$_2$ | $a$ = 2.801Å $c$ = 12.806Å | K(2b) | 0.500 | 0.500 | 0.000 | +0.489 | 9.24 |
| | | K(4e) | 0.500 | 0.500 | 0.798 | +0.405 | 9.12 |
| | | Cl(4e) | 0.000 | 0.000 | 0.897 | -0.655 | 11.39 |
| $I4/mmm$-K$_4$Cl$_3$ | $a$ = 2.791Å $c$ = 18.190Å | K(4e) | 0.000 | 0.000 | 0.784 | +0.413 | 9.13 |
| | | K(4e) | 0.000 | 0.000 | 0.072 | +0.449 | 9.17 |
| | | Cl(2b) | 0.000 | 0.000 | 0.500 | -0.673 | 11.66 |
| | | Cl(4e) | 0.000 | 0.000 | 0.646 | -0.621 | 11.31 |
| $I4/mmm$-K$_5$Cl$_4$ | $a$ = 2.789Å $c$ = 23.531Å | K(2a) | 0.000 | 0.000 | 0.000 | +0.569 | 9.19 |
| | | K(4e) | 0.500 | 0.500 | 0.612 | +0.563 | 9.13 |
| | | K(4e) | 0.500 | 0.500 | 0.276 | +0.410 | 9.12 |
| | | Cl(4e) | 0.000 | 0.000 | 0.669 | -0.622 | 11.32 |
| | | Cl(4e) | 0.000 | 0.000 | 0.557 | -0.636 | 11.58 |

(a)

(b)

(c)

**Figure 1. Stability of new potassium chlorides:** (a) Pressure-composition phase diagram of the K-Cl system. (b-c) Convex hull diagrams for the K-Cl system at selected pressures. Solid circles represent stable compounds; open circles - metastable ones.

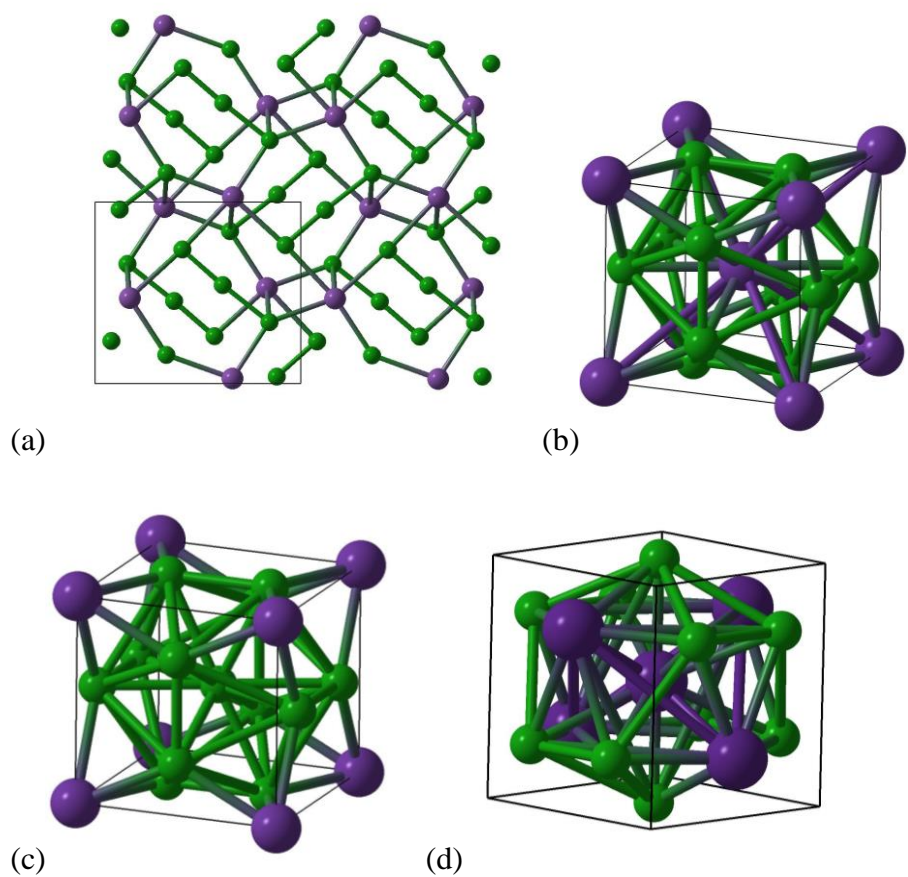

(a)  (b)

(c)  (d)

**Figure 2. Crystal structures of** (a) *Pnma*-KCl$_3$ at 20 GPa (b) *Pm*3*n*-KCl$_3$ at 240 GPa (c) *Pm*3-KCl$_7$ at 240 GPa, (d) K$_3$Cl$_5$-$P\bar{4}m2$ at 240 GPa. Large gray spheres – K atoms, small green spheres – Cl atoms.

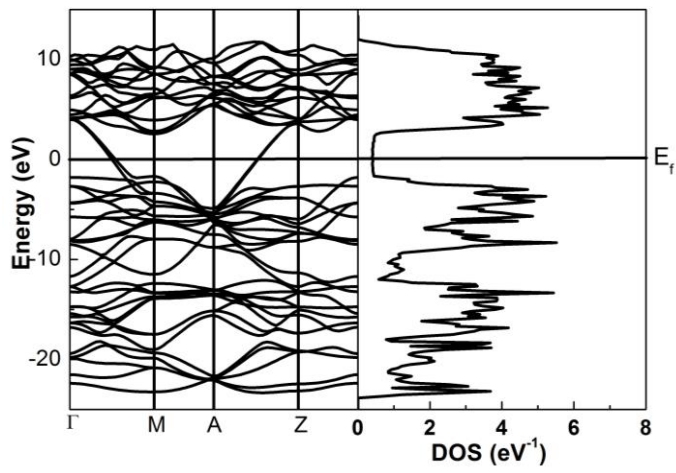
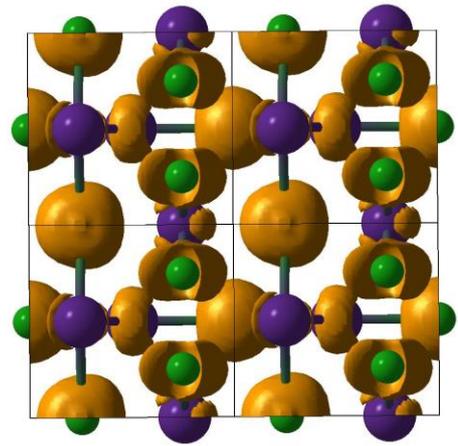

(a) (b)

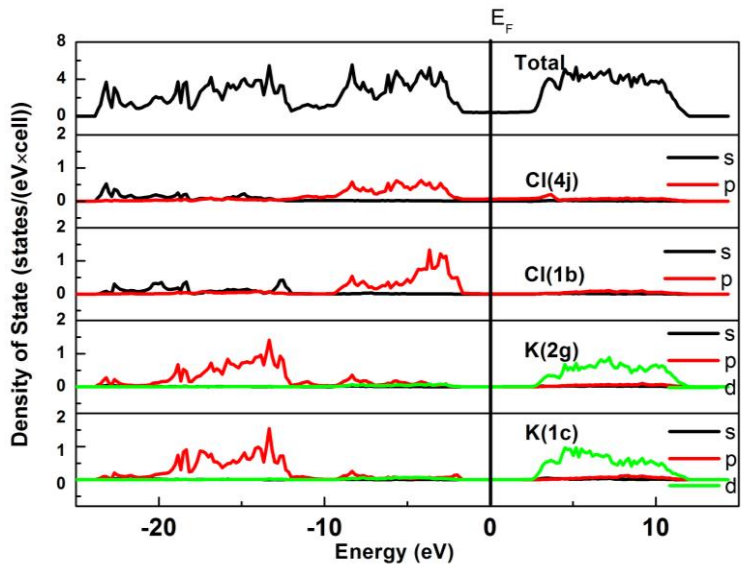

(c)

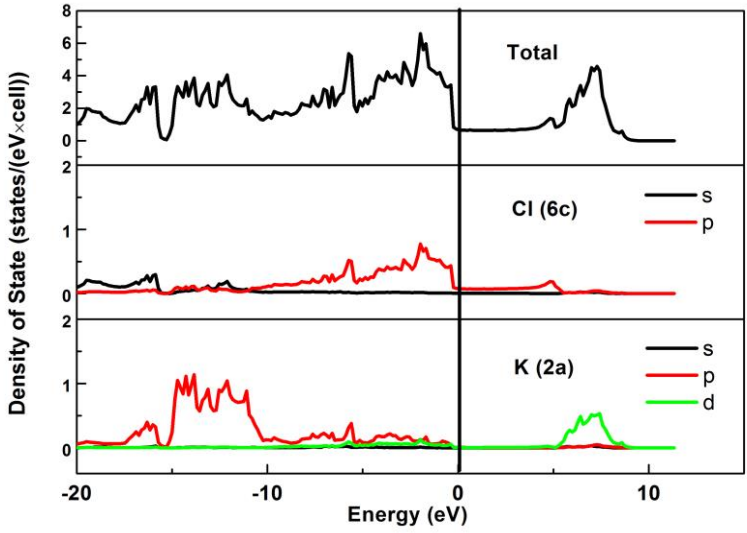

(d)

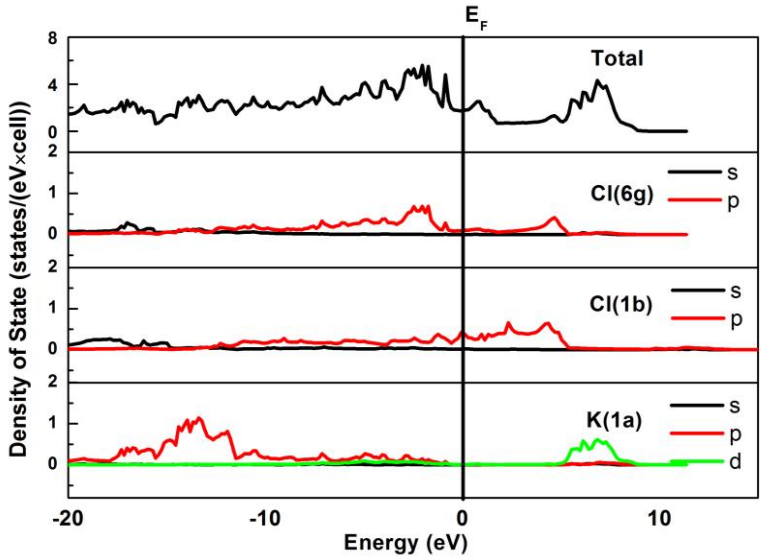

(e)

**Figure 3. Electronic structure:** (a) band structure and electronic density of states of $P\bar{4}m2$-$K_3Cl_5$ at 240 GPa, (b) electron localization function of $P\bar{4}m2$-$K_3Cl_5$ at 240 GPa with isosurface ELF=0.77. (c) total and atom-projected densities of states of $P\bar{4}m2$-$K_3Cl_5$, (d) total and atom-projected densities of states of $Pm3n$-$KCl_3$, (e) total and atom-projected densities of states of $Pm3$-$KCl_7$ at 240 GPa.

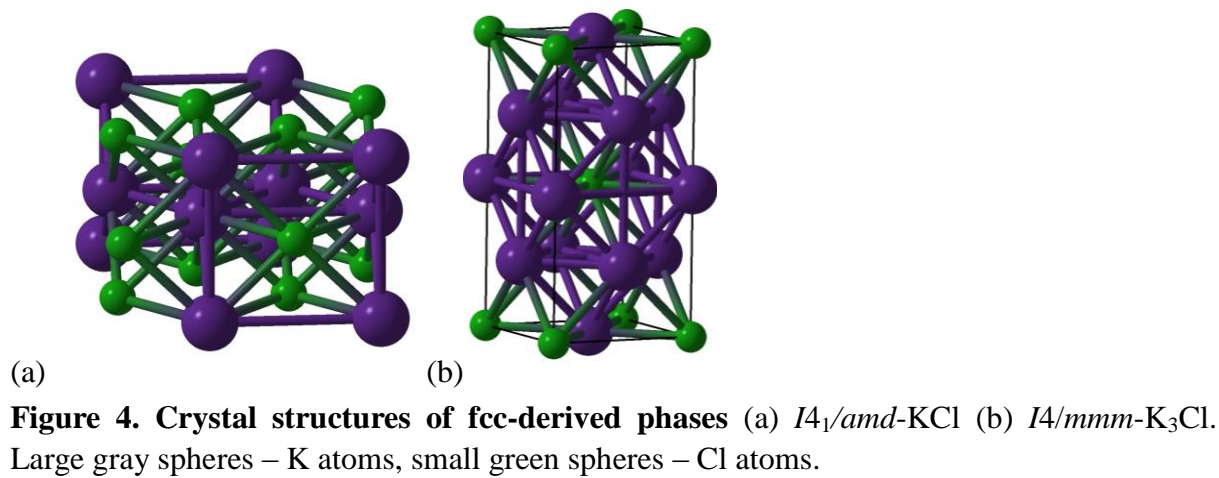

**Figure 4. Crystal structures of fcc-derived phases** (a) $I4_1/amd$-KCl (b) $I4/mmm$-$K_3Cl$. Large gray spheres – K atoms, small green spheres – Cl atoms.

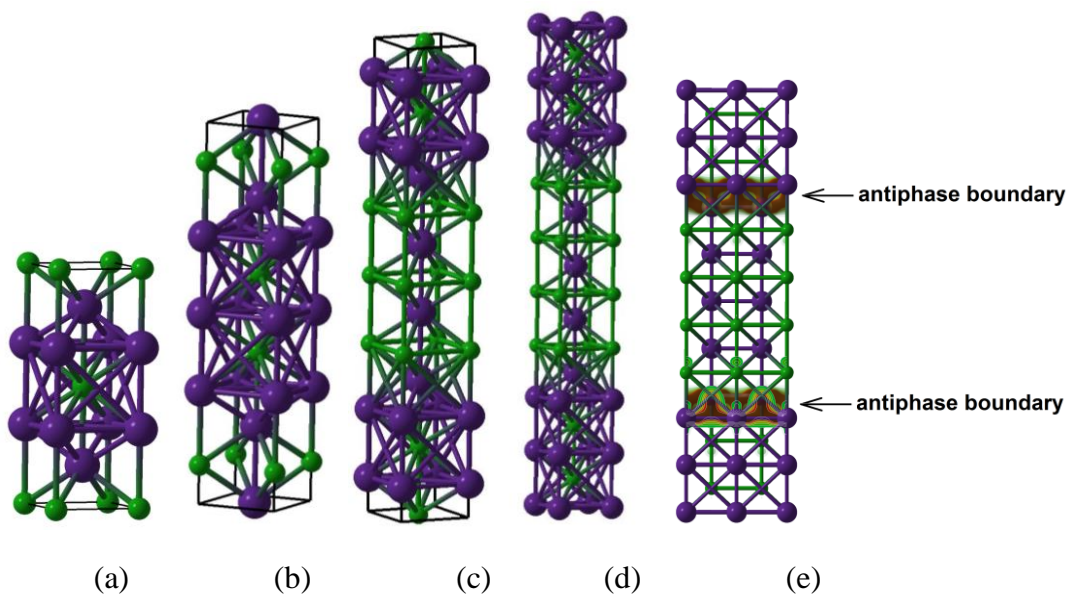

**Figure 5. Crystal structures of $K_{n+1}Cl_n$ homologs** (a) $I4/mmm$-$K_2Cl$, (b) $I4/mmm$-$K_3Cl_2$, (c) $I4/mmm$-$K_4Cl_3$, (d) $I4/mmm$-$K_5Cl_4$ (e) Spatial distribution of electrons (shown by isosurfaces and density contours) at the Fermi level in $I4/mmm$-$K_5Cl_4$, showing clear 2D-metallic character. Large gray spheres – K atoms, small green spheres – Cl atoms.

**Supplementary Information**

Figs. S1 to S3, Table S1 to S2.

# Supplementary Information for

Stability of numerous novel potassium chlorides at high pressure

Weiwei Zhang[1, 2, *], Artem R. Oganov[2-4, *]

*Corresponding author. E-mail: zwwjennifer@gmail.com (W.Z.); artem.oganov@sunysb.edu (A.R.O.)

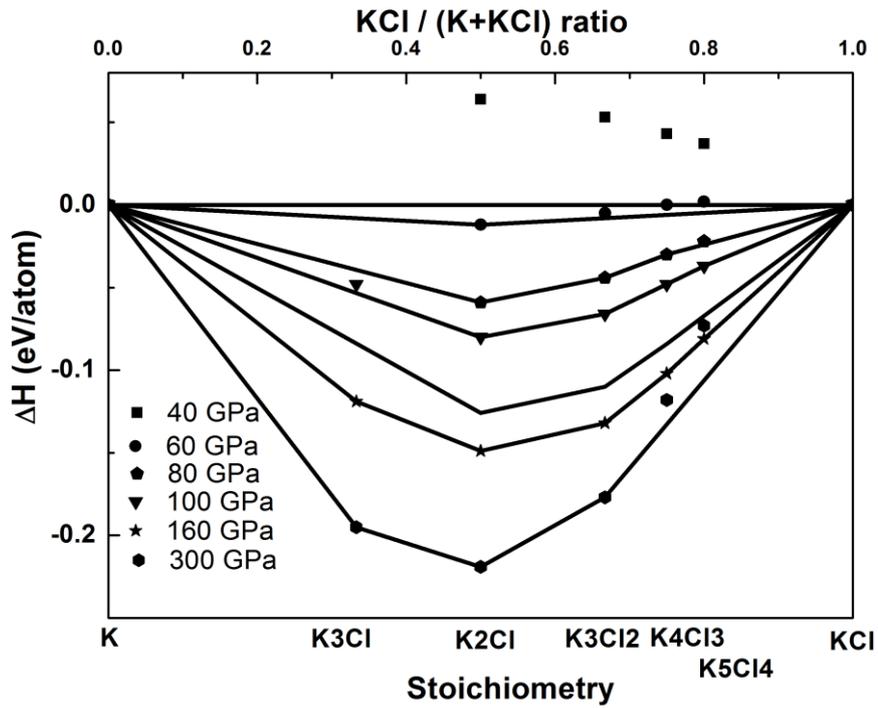

(a)

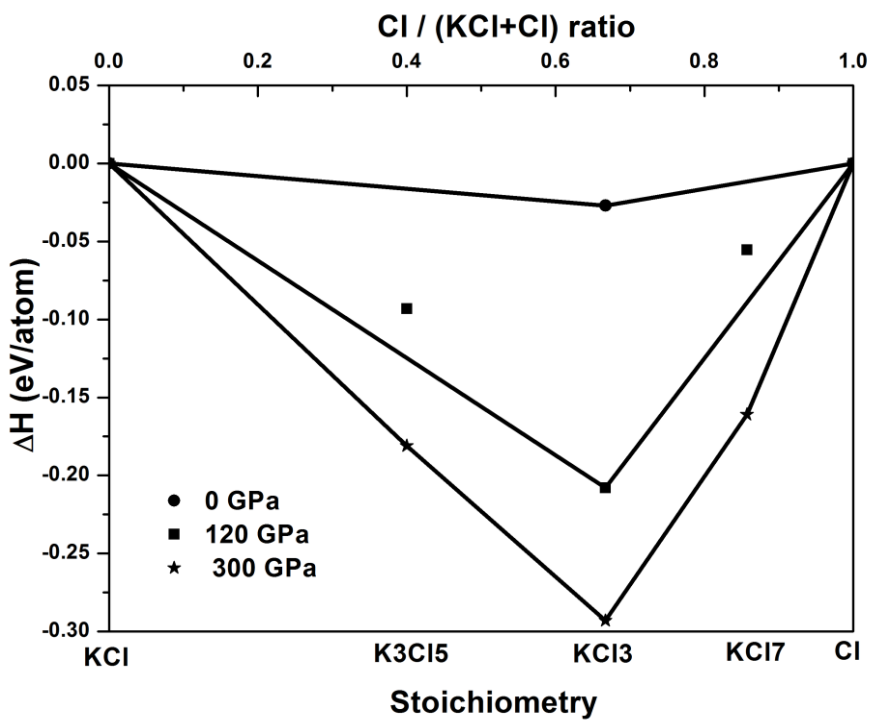

(b)

Figure S1. Convex hull diagrams for (a) K-KCl system and (b) KCl-Cl system.

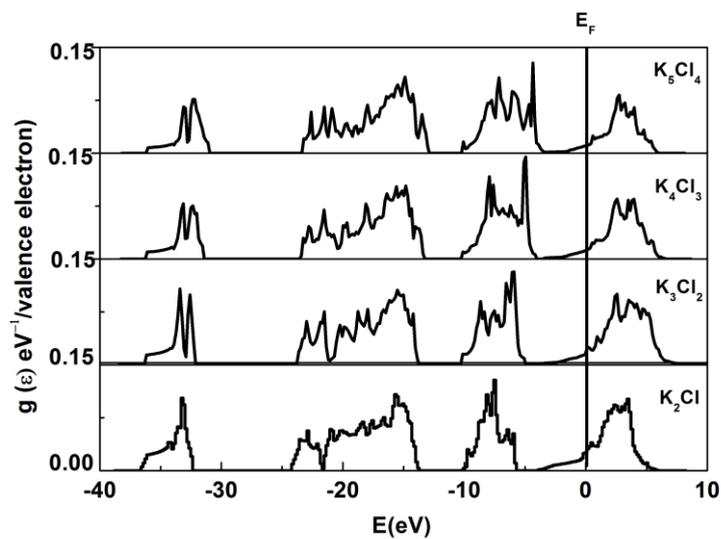

Figure S2. The electronic density of states of layered $K_{n+1}Cl_n$ homologs at 200 GPa.

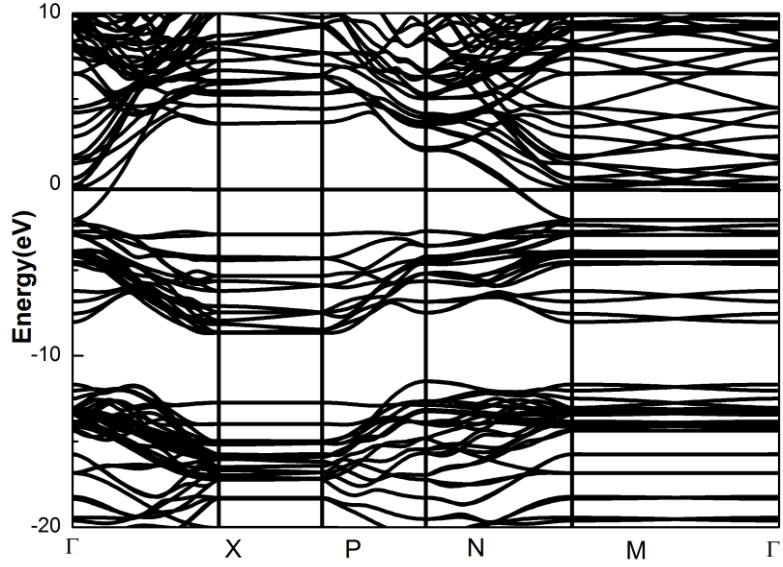

Figure S3. Band structure of *I4/mmm*-K$_5$Cl$_4$ at 200GPa.

Table S1. Structure information and corresponding stability pressure regions for new compounds

| Class of phases | Composition | Pressure range of stability | Space group | Cell parameters and atomic coordinates | |
|---|---|---|---|---|---|
| Insulating KCl | post-B2-KCl | 201-300 | $I4_1/amd$ | $a$ = 3.368Å<br>$c$ = 6.955Å<br>(220 GPa) | K1  4a(0.0, 0.25, 0.875)<br>Cl1  4b(0.0, 0.75, 0.625) |
| Insulating Cl-rich phase | Pnma-KCl$_3$ | 0-34 | *Pnma* | $a$=8.149Å<br>$b$=4.910Å<br>$c$=7.276Å (20 GPa) | K1  4c (x, 0.25, z)<br>     x=0.166, z= 0.471<br>Cl1  4c (x, 0.25, z)<br>     x = 0.922, z = 0.763<br>Cl2  4c (x, 0.25, z)<br>     x = 0.618, z = 0.543<br>Cl3  4c (x, 0.25, z)<br>     x = 0.845, z = 0.341 |
| Metallic Cl-rich phases | KCl$_3$ | 34-300 | $Pm3n$ | $a$ = 4.169 Å (240 GPa) | K   2a (0.0, 0.0, 0.0)<br>Cl  6c (0.0, 0.5, 0.25) |
| | KCl$_7$ | 225-300 | $Pm3$ | $a$ = 4.123 Å (240 GPa) | K   1a (0.0, 0.0, 0.0)<br>Cl1  1b (0.5, 0.5, 0.5)<br>Cl2  6g (x, 0.5, 0.0)<br>     x=0.745 |
| | K$_3$Cl$_5$ | 142-300 | $P\bar{4}m2$ | $a$ =4.136Å<br>$c$ = 4.361Å<br>(240 GPa) | K1  1c (0.5, 0.5 ,0.5)<br>K2  2g (0.5, 0.0, -z )<br>    z=-0.245<br>Cl1 1b (0.5, 0.5, 0.0)<br>Cl2 4j (x, 0.0, z)<br>    x=0.252, z=0.747 |
| Metallic K$_{n+1}$Cl$_n$, B2-layered superstructures | K$_2$Cl | 56-300 | $I4/mmm$ | $a$= 2.749Å<br>$c$ = 7.587Å<br>(200 GPa) | K1  4e (0.0,0.0,z)<br>    z=0.330<br>Cl  2a (0.0,0.0,0.0) |
| | K$_3$Cl$_2$ | 80-300 | $I4/mmm$ | $a$ = 2.801Å<br>$c$ = 12.806Å<br>(200 Gpa) | K1  2b (0.5,0.5,0.0)<br>K2  4e ( 0.5,0.5,z)<br>    z=0.798<br>Cl  4e (0.0,0.0,z )<br>    z=0.897 |
| | K$_4$Cl$_3$ | 62-264 | $I4/mmm$ | $a$ = 2.791Å<br>$c$ =18.190Å<br>(200 GPa) | K1  4e (0.0, 0.0, z)<br>    z=0.072<br>K2  4e (0.0, 0.0, z)<br>    z=0.784<br>Cl1 2b(0.0, 0.0, 0.5)<br>Cl2 4e (0.0, 0.0, z)<br>    z= 0.646 |
| | K$_5$Cl$_4$ | 100-255 | $I4/mmm$ | $a$ =2.789Å<br>$c$ =23.531Å<br>(200 GPa) | K1  2a (0.5, 0.5, 0.5)<br>K2  4e(0.5, 0.5, z)<br>    z = 0.612<br>K3  4e(0.5, 0.5, z)<br>    z=0.276<br>Cl1 4e (0.0,0.0,z)<br>    z=0.669<br>Cl2 4e (0.0,0.0,z)<br>    z=0.557 |
| Metallic fcc-layered superstructure | K$_3$Cl | 149-300 | $I4/mmm$ | $a$=3.427Å<br>$c$=6.714 Å<br>(200 Gpa) | K1  2b(0.5,0.5,0.0)<br>K2  4d (0.5,0.0,0.25)<br>Cl  2a(0.0,0.0,0.0) |

**Table S2.** Structures of A15-type (*Pm3n*) KCl$_3$ and NaCl$_3$ at 200 GPa, and the corresponding atomic Bader charges (Q) and volumes (V).

| | | | | | | |
|---|---|---|---|---|---|---|
| *Pm3n*-KCl$_3$ | K(2a) | 0.000 | 0.000 | 0.000 | +0.555 | 8.29 |
| | Cl(6c) | 0.000 | 0.500 | 0.250 | -0.185 | 10.02 |
| *Pm3n*-NaCl$_3$ | Na(2a) | 0.000 | 0.000 | 0.000 | +0.823 | 4.16 |
| | Cl(6d) | 0.250 | 0.500 | 0.000 | -0.275 | 9.90 |